\documentclass[conference]{IEEEtran}
\IEEEoverridecommandlockouts

\usepackage{setspace}
\usepackage[T1]{fontenc}
\usepackage{color,array,amsthm}
\usepackage{graphicx}
\usepackage{cite}
\usepackage{amsmath,amssymb,amsfonts}
\usepackage{textcomp}
\usepackage{xcolor}
\usepackage{colortbl}
\usepackage{url}
\usepackage{multirow}
\usepackage{graphicx}
\usepackage{wrapfig}
\usepackage{latexsym}
\usepackage{algorithm}
\usepackage{algpseudocode}
\usepackage{subcaption}
\usepackage[top=0.70in, bottom=1.05in, left=0.64in, right=0.64in]{geometry}
\def\BibTeX{{\rm B\kern-.05em{\sc i\kern-.025em b}\kern-.08em
    T\kern-.1667em\lower.7ex\hbox{E}\kern-.125emX}}
\begin{document}


\title{Green Traffic Engineering for Satellite Networks Using Segment Routing Flexible Algorithm\\

}
\author{
\IEEEauthorblockN{
Jintao Liang\IEEEauthorrefmark{1},
Pablo G. Madoery\IEEEauthorrefmark{1},
Chung-Horng Lung\IEEEauthorrefmark{2},
Halim Yanikomeroglu\IEEEauthorrefmark{1},
Gunes Karabulut Kurt\IEEEauthorrefmark{1}\IEEEauthorrefmark{3}
}
\vspace*{0.3cm}
\IEEEauthorblockA{\IEEEauthorrefmark{1} Non-Terrestrial Networks (NTN) Lab, Department of Systems and Computer Engineering, Carleton University, Canada}
\IEEEauthorblockA{\IEEEauthorrefmark{2}Department of Systems and Computer Engineering, Carleton University, Canada}
\IEEEauthorblockA{\IEEEauthorrefmark{3} Poly-Grames Research Center, Department of Electrical Engineering, Polytechnique Montréal, Montréal, Canada}
}
\maketitle

\begin{abstract}
Large-scale low-Earth-orbit (LEO) constellations demand routing that simultaneously minimizes energy, guarantees delivery under congestion, and meets latency requirements for time-critical flows. We present a segment routing over IPv6 (SRv6) flexible algorithm (Flex-Algo) framework that consists of three logical slices: an energy-efficient slice (Algo 130), a high-reliability slice (Algo 129), and a latency-sensitive slice (Algo 128). The framework provides a unified mixed-integer linear program (MILP) that combines satellite CPU power, packet delivery rate (PDR), and end-to-end latency into a single objective, allowing a lightweight software-defined network (SDN) controller to steer traffic from the source node. Emulation of Telesat's Lightspeed constellation shows that, compared with different routing schemes, the proposed design reduces the average CPU usage by 73\%, maintains a PDR above 91\% during traffic bursts, and decreases urgent flow delay by 18 ms between Ottawa and Vancouver. The results confirm Flex-Algo’s value as a slice-based traffic engineering (TE) tool for resource-constrained satellite networks.
\end{abstract}


\vspace{-0.05cm}
\section{Introduction}

Non-terrestrial networks (NTNs), particularly low-Earth-orbit (LEO) satellite constellations, have emerged as a transformative approach to global connectivity \cite{future_space_networks}. Unlike traditional geostationary systems, LEO satellite constellations, such as Telesat’s Lightspeed \cite{lightspeed_fcc}, offer lower latency and higher spatial diversity than ever before. However, the scale and dynamic topology of these modern constellations expose the limitations of legacy architectures, demanding more flexible network management, seamless integration with terrestrial technologies, and advanced control mechanisms to dynamically steer traffic and optimize performance across diverse scenarios \cite{evolution_ntn}.

Segment routing (SR) simplifies network routing by encoding the packet’s path directly within the packet via a sequence of segment identifiers (SIDs) \cite{rfc8402}. This approach streamlines the control plane, reduces the configuration complexity, and enables more flexible and efficient traffic management \cite{sr architecture}. When implemented over IPv6 (SRv6), the SR header (SRH) encapsulates a list of IPv6 addresses as segments \cite{rfc8754}. In LEO satellite constellations, using SRv6 for green traffic engineering (TE) can significantly reduce onboard CPU load, thereby reducing energy consumption without sacrificing other key metrics \cite{green_srv6}. When integrated with software-defined network (SDN), SRv6 enables dynamic control over the network, supporting green TE for energy-constrained satellite constellations.

Flexible algorithm (Flex-Algo) represents a pivotal innovation in SR, extending interior gateway protocols (IGP) to compute constraint-based paths within the control plane \cite{cisco_flex_algo}. Flex-Algo can create multiple virtual topologies based on different real-time metrics, such as latency and energy consumption, and dynamically select paths at the source node once the SID stack is assigned. The inherent flexibility of Flex-Algo simplifies network management, enhances scalability, and supports intent-based networking frameworks \cite{huawei_flex_algo}, making it particularly suitable for satellite networks that must rapidly adapt to link failures, load fluctuations, and energy constraints.

\begin{figure}[b]
    \vspace{-0.6cm}
    \centering
    \includegraphics[width=0.85\linewidth]{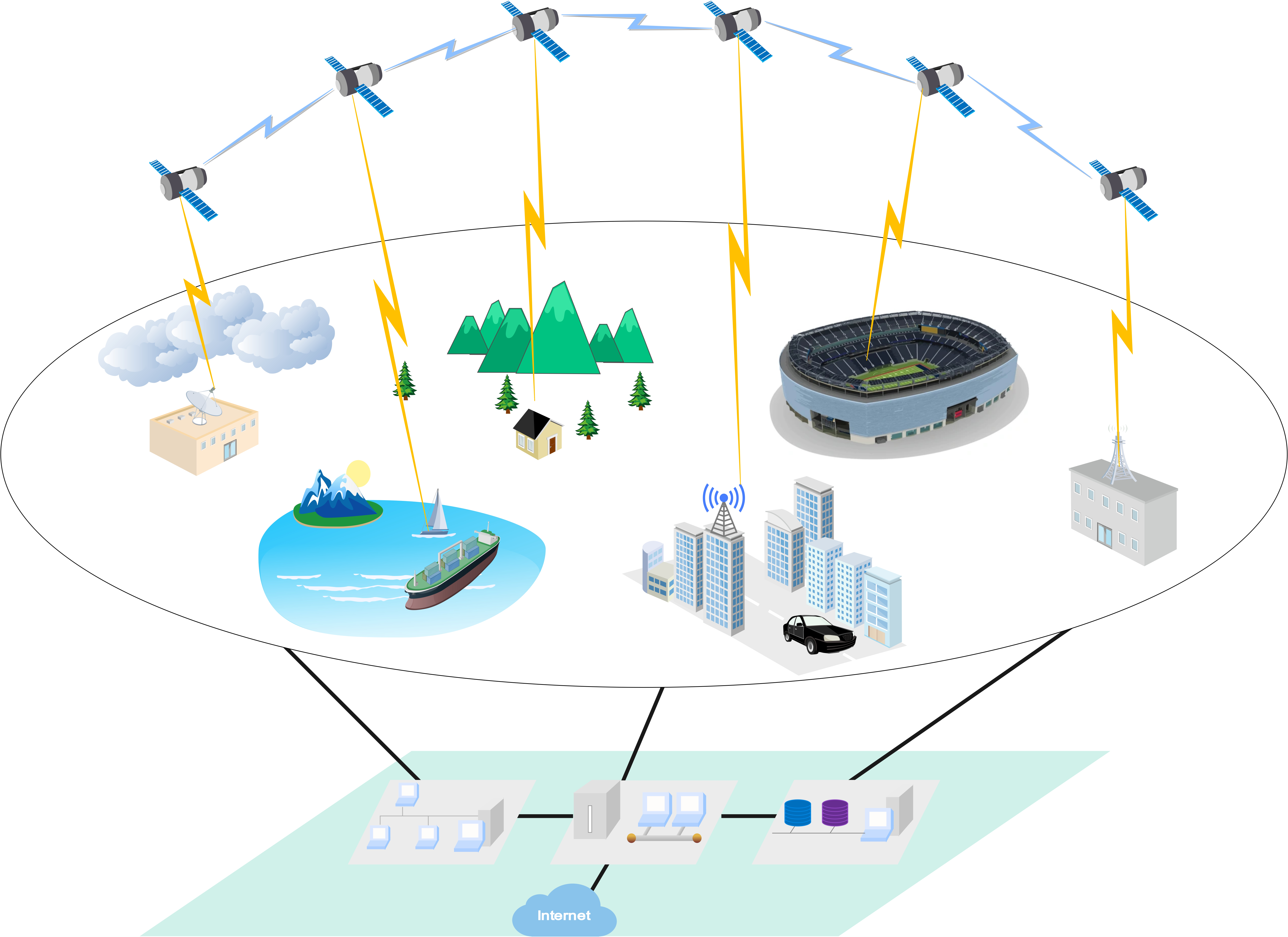}
    \caption{System architecture of LEO satellite network.}
    \label{system_model_fig}
\end{figure}

This paper proposes a comprehensive framework that integrates three SRv6 Flex-Algos to achieve green TE in LEO satellite networks. We simulate Telesat's Lightspeed constellation, formulate the optimization problem, and evaluate key performance metrics. By evaluating three slices—energy (Algo 130), reliability (Algo 129), and latency (Algo 128)—against the baseline routing stacks (IPv6 open shortest path first (OSPF), plain SRv6, and our prior green SRv6 \cite{green_srv6}), clear patterns emerge. Under default traffic, Flex-Algo matches Green SRv6’s low CPU usage. During high-demand bursts, it selects Algo 129 to preserve a high packet delivery rate (PDR) with only minimal power and latency increases. For time-critical flows, it switches to Algo 128 to cut end-to-end delay while keeping power low. Overall, Flex-Algo offers the best energy–reliability–latency trade-off for resource-constrained LEO constellations. The contributions of this work are as follows: 
\begin{itemize}
  \item An SDN-based LEO satellite network architecture is designed with three SRv6 Flex-Algo slices, each focusing on a specific objective: energy efficiency, high reliability, and low latency. 
  \item A unified mixed-integer linear program (MILP) is formulated to jointly optimize energy consumption, reliability, and end-to-end latency within a single objective, subject to practical and realistic constraints.
  \item An extensive Mininet-based emulation in the Lightspeed constellation using Ryu controller demonstrates that the proposed Flex-Algo slices achieve high PDR under high demands, maintain low latency for urgent flows, and sustain CPU usage comparable to dedicated green TE.
\end{itemize}

The remainder of the paper is organized as follows: Section II covers related work and motivation. Section III presents the system model, the Flex-Algo configurations, and the mathematical formulation. Section IV details the simulation setup and analyzes performance results. Finally, Section V concludes the paper and outlines directions for future research.

\vspace{-0.05cm}
\section{Related Work}

Several studies discuss green TE and SR implementation in satellite networks, and Flex-Algo as discussed in \cite{green_traffic_engineering, energy_eff_sat, green_sat, sr te, sr_and_green_sdn, green_srbackbone, flexalgo_user_path, flexalgo_evaluation, cisco_flex_algo, huawei_flex_algo, green_srv6}. 

In \cite{green_traffic_engineering}, green TE is implemented in SDN with SR by introducing low-power idle modes and control-plane power-state signaling. This approach reduces energy consumption while preserving performance under dynamic loads. In \cite{energy_eff_sat}, the authors examine energy savings across satellite systems, highlighting adaptive power control and resource allocation via cross-layer optimization to achieve green TE. In \cite{green_sat}, the authors propose using green SR to minimize power draw and path stretch, select low-power links, and enable sleep modes to extend satellite battery lifetimes.

In \cite{sr te}, the work presents an SDN-SR integrated TE architecture with an open-source implementation and a practical flow allocation heuristic. In \cite{sr_and_green_sdn}, integrating SR with SDN enables green TE by leveraging SR segments and energy-aware routing. Flexible SR policies that consider energy usage and resilience enable balanced route optimization, improving both energy efficiency and fault tolerance. In \cite{green_srbackbone}, the authors propose a SR–based TE algorithm that consolidates traffic onto minimal links to power down hardware during idle periods while enforcing maximum link utilization constraints. 

In \cite{cisco_flex_algo}, Cisco outlines the extensions to support Flex-Algo. It explains the prerequisites, configuration steps, and provides practical examples of actual setup. In \cite{huawei_flex_algo}, Huawei highlights that Flex-Algo enables independent logical topologies, supports flexible metric selection for diverse services, and integrates with SR for efficient and resilient routing. In \cite{flexalgo_user_path}, the authors achieve user-driven path control in SR, implementing Flex-Algo across three scenarios and building an automated traffic-steering tool driven by user metrics and constraints. In \cite{flexalgo_evaluation}, the paper evaluates Flex-Algo for quality-of-service-driven TE in SR with two topologies—latency-sensitive and best effort—and compares against single-slice baselines.

Building on prior research, numerous works have studied green TE and SR in various network contexts. In our previous work \cite{green_srv6}, we proposed a green TE using SRv6 and introduced a comparative analysis. The results showed that the proposed green TE significantly reduces onboard CPU usage without compromising other key metrics. Here, we extend to Flex-Algo: three SRv6 slices—energy, reliability, and latency—driven by a unified SDN control plane and a MILP. This real-time framework directs high demands to the reliability slice and urgent flows to the latency slice, ensuring that different performance requirements are satisfied for various types of traffic.

To the best of our knowledge, this is the first study to investigate SRv6 Flex-Algo in MILP that jointly optimizes energy, reliability, and latency within a LEO satellite constellation. Our work also compares dynamic Flex-Algo against Green TE and legacy routing schemes under diverse traffic conditions.

\vspace{-0.05cm}
\section{System Model and Methodology}
In this section, we present the system model and methodology used in this work, including the simulation of the Lightspeed satellite constellation, the configurations of Flex-Algo slices, and the MILP formulation. 

\subsection{System Model}

\begin{table*}[t]
\centering
\caption{Flex-Algos configurations}
\label{flex_algo_table}
\begin{tabular}{|c|c|c|c|}
\hline
\textbf{Flex-Algo (ID)} & \textbf{Metric type} & \textbf{Link filter Rules} & \textbf{Use case} \\ \hline
130 – Energy efficiency & Energy \(p_{ij}\) & Default mode & Green mode \\ \hline
129 – High reliability   & Utilization \(u_{ij}\) and \(PDR_{ij}\) & Exclude \texttt{low-rely} links & Hot spot events  \\ \hline
128 – Low latency   & Delay \(d_{ij}\) & Low latency required  & Time sensitive traffic \\ \hline
\end{tabular}
\vspace{-0.5cm}
\end{table*}

As shown in Fig. \ref{system_model_fig}, the system adopts a three-layer LEO satellite architecture: yellow lines denote uplinks/downlinks, blue lines denote inter-satellite links (ISLs), and black lines connect ground stations and SDN controllers. As an SDN-based network, the satellites forward data packets between ground stations under the control of the centralized SDN controller, which dynamically adjusts routing paths based on real-time network topology and traffic conditions. Satellites establish communication links with various types of ground terminals, such as event hot spots (e.g., stadiums with high user density and traffic volume) and institutions (e.g., stock exchanges with extreme latency sensitivity).

We simulate Telesat’s Lightspeed LEO satellite constellation, a well-known network designed to provide global broadband connectivity that has not yet been fully deployed. The constellation comprises 78 polar-orbit satellites at 1,015 km altitude and 120 inclined-orbit satellites at 1,325 km.

In this work, satellites are assigned IPv6 addresses and configured to support SRv6. Their flow tables are programmed using the OpenFlow 1.3 protocol to match active SIDs and perform the required forwarding actions. When a packet arrives, the satellite decrements the SRH, updates the destination to the next SID, and forwards the packet. We use iperf3 to generate traffic and evaluate SRv6 under varying traffic loads. To forward along Flex-Algo paths, satellites install IPv6 routes with the algorithm’s prefix-SID; only packets with the advertised prefix-SIDs follow that algorithm. An on-demand next-hop template automatically instantiates the SR policy and programs the per-slice prefix-SID. The Flex-Algo workflow comprises four steps: algorithm definition, algorithm advertisement, topology generation, and path calculation.

\subsection{Flex-Algo Slices Configurations}

\begin{figure}
    \centering
    \includegraphics[width=0.85\linewidth]{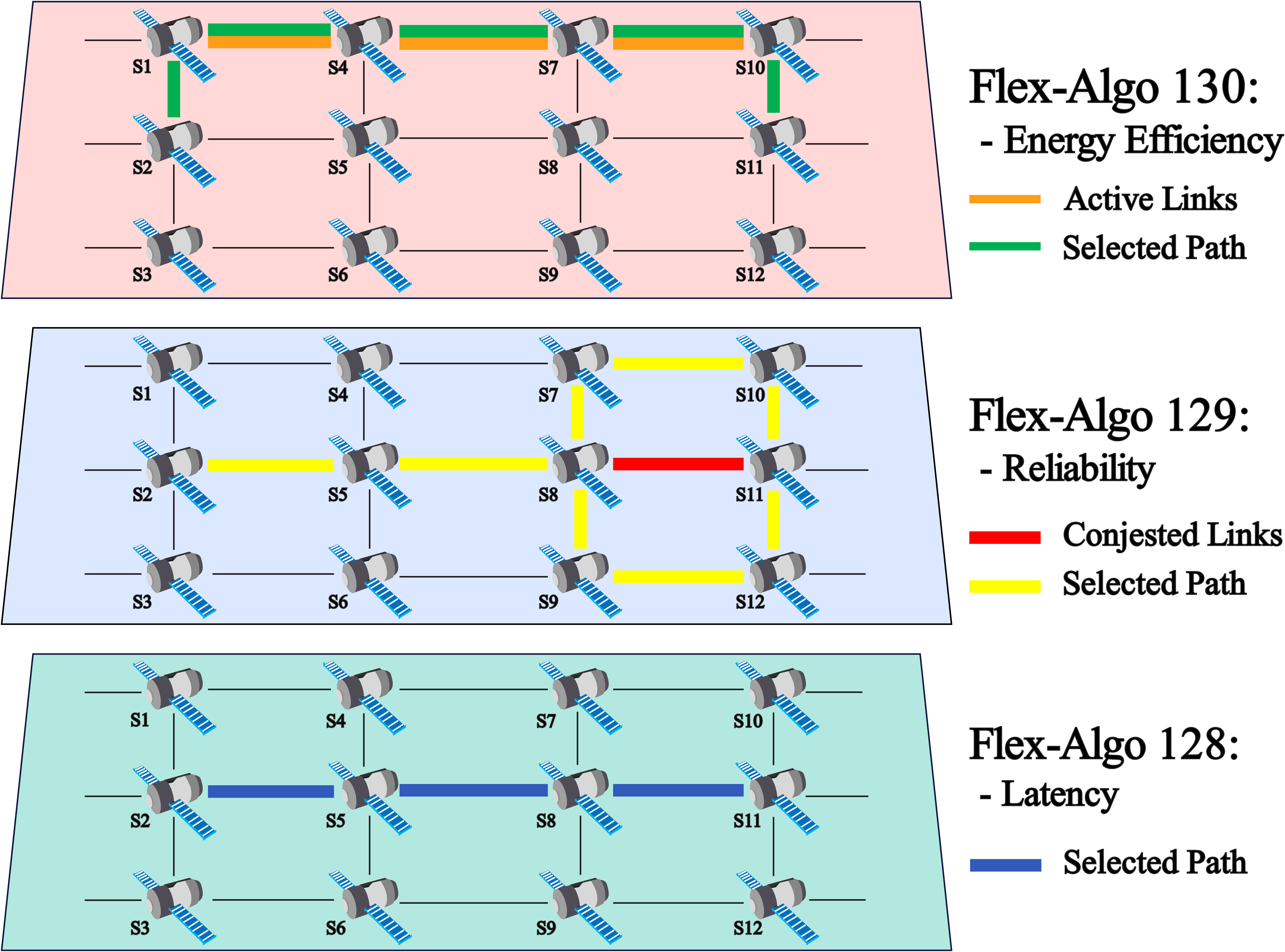}
    \caption{Flex-Algo slices demonstration.}
    \label{flex_algo_fig}
     \vspace{-0.6cm}
\end{figure}

This subsection specifies the SRv6 Flex-Algo slices used in our study. Flex-Algo computes multiple parallel shortest-path trees from a single link-state database by varying metrics and link filters. By steering traffic to the SR policy or locator associated with one specific algorithm, the source node selects the slice that best matches the requirements. We define three configurations—energy, reliability, and latency—summarized in Table \ref{flex_algo_table}. The detailed per-slice settings are as follows.

Flex-Algo 130 for energy efficiency: 
\begin{itemize}
    \item Definition: The calculation type is Dijkstra’s shortest path first (SPF) algorithm. The metric type is TE metric, which is represented by the satellites’ onboard CPU usage to indicate energy consumption \(p_i\) for satellite \(i\). The routing constraints include both peak power limitations and total latency constraints to ensure efficiency and reliability.

    \item Implementation: Satellites advertise an SRv6 locator using algorithm 130 to maintain green TE.

    \item Use case: This configuration serves as the default mode under regular traffic to minimize the total energy consumption across the satellite constellation.
\end{itemize}

Flex-Algo 129 for high reliability:
\begin{itemize}
    \item Definition: The calculation type remains SPF. The metric-type is the TE metric, computed as \(R_{ij}=PDR_{ij}\bigl(1-u_{ij}\bigr)\) to estimate the reliability and congestion state of link \((i,j)\), where \(PDR_{ij}\) is the recent PDR and \(u_{ij}\) the current utilization. An extra routing constraint affinity is added besides the peak power and total latency limitations.

    \item Implementation: All satellites advertise an SRv6 locator with algorithm 129. Links whose $R$ are low are tagged with the affinity bit \texttt{low-rely}. Those links are excluded for SPF computation, ensuring that only highly reliable links enter the logical topology.

    \item Use case: This algorithm is activated during high-demand scenarios, such as hot spot events like football matches or under severe weather. It prioritizes routing paths that maximize delivery success and reliability, even if this results in longer paths or higher energy consumption. 
\end{itemize}

Flex-Algo 128 for low latency: 
\begin{itemize}
    \item Definition: The calculation type remains SPF, with the metric type defined as the propagation delay \(d_{ij}\) of link \((i,j)\). Additionally, node delay \(T_{node}\) per satellite is also considered to include the handover of ISLs\cite{fso_latency_power}. Peak power consumption is also included as a routing constraint to ensure sustainable operation.

    \item Implementation: Satellites advertise an SRv6 locator under algorithm 128 to guarantee the lowest latency transmission path across the network.

    \item Use case: This configuration is selected for time-critical traffic, such as stock market transactions and interactive control loops, where even microseconds of delay can significantly affect mission success or financial outcomes.
\end{itemize}

Fig. \ref{flex_algo_fig} illustrates the topology produced by each Flex-Algo on the same $3\times4$ satellite example mesh. The pink pane (Algo 130) selects the least-energy route (green) over already-active links (orange). The blue pane (Algo 129) prunes congested links (red), so the algorithm chooses a longer but safer path (yellow). The green pane (Algo 128) picks the minimum-hop path with the lowest latency (blue). Since all three trees are precomputed, switching slices only requires updating the source SR policy. For simplicity, Doppler effect in space and its impacts on reliability/latency are not considered in this work.

\subsection{Mathematical Formulation}
We aim to optimize network performance based on three Flex-Algo slices by minimizing total satellite energy consumption (indicated by CPU power consumption), path reliability prediction penalty, and end-to-end propagation with node delay. The MILP enforces flow conservation, link capacity, node degree, and peak power constraints to ensure feasible, energy-aware, and latency-sensitive routing across the network.

\subsubsection{Inputs}

\begin{itemize}
    \item $S$ — set of source ground stations; $s^{k}$ and $d^{k}$ are the source and destination of connection $k\in K$;
    \item $K$ — set of traffic connections to be routed;
    \item \(d_{ij}\) — propagation delay (ms) of link $(i,j)$;
    \item \(C_{ij}\) — capacity (Mbps) of link $(i,j)$;
    \item \(b_{k}\) — bandwidth demand (Mbps) of connection $k$;
    \item $p_i$ — CPU power usage of satellite $i$;
    \item \(P_{{lim}}\) — peak CPU power usage limit per satellite;
    \item \(T_{{node}}\) — node delay (ms), here fixed to $10$ ms;
    \item \(R_{ij}\) — composite link reliability score ($0<R_{ij}\le 1$);
    \item $\alpha,\,\beta,\,\gamma$ — non-negative weights for energy, reliability, and latency, respectively.
\end{itemize}

\subsubsection{Variables}

\begin{itemize}
    \item $x_{ij}^{k}$ — 1 if link $(i,j)$ is in the path of connection $k$;
    \item $y_{i}^{k}$ — 1 if satellite $i$ lies on the path of connection $k$;
    \item $z_{i}$ — 1 if satellite $i$ is activated  by any connection (\,$z_i=1$ if $\sum_{k}y_{i}^{k}\ge 1$\,), and 0 otherwise.
\end{itemize}

\subsubsection{Objective}
The objective function can be expressed as
\begin{equation}
\label{eq_obj}
\begin{split}
\min\;&
\underbrace{\alpha\sum_{i} p_{i}\,z_{i}}_{\text{total CPU power usage}}
+
\underbrace{\beta
  \sum_{(i,j)}\bigl[-\log R_{ij}\bigr]
        \sum_{k}x_{ij}^{k}
}_{\text{path reliability penalty}}
\\[4pt]
&+
\underbrace{\gamma
  \Bigl(
        \sum_{(i,j)}d_{ij}\sum_{k}x_{ij}^{k}
        + T_{\mathrm{node}}\sum_{i}\sum_{k}y_{i}^{k}
  \Bigr)}_{\text{end-to-end latency}}
  \text{.}\; 
\end{split}
\vspace{-0.3cm}
\end{equation}

The first term minimizes the total satellite CPU power consumption across the network. The second term minimizes the negative logarithm of the estimated reliability score (i.e., maximizing each packet’s end-to-end success probability). The third term sums the propagation and node delay over all paths, thereby identifying the fastest routes.

\subsubsection{Constraints}

Flow-conservation constraints
\begin{equation}
\vspace{-0.1cm}
\label{eq_flow}
\sum_{j}x_{hj}^{k}-\sum_{i}x_{ih}^{k}=
\begin{cases}
  1, & h=s^{k},\\
 -1, & h=d^{k},\\
  0, & \text{otherwise}, 
\end{cases}
\qquad\forall h,\;k \text{.}\; 
\vspace{-0.1cm}
\end{equation}
All traffic that leaves a source must reach its destination, with no creation or loss of flow at intermediate satellites.

Node-degree constraints
\begin{equation}
\label{eq_degree}
\sum_{i}x_{ih}^{k}+\sum_{j}x_{hj}^{k}\le 4,
\qquad\forall h,\;k \text{.}\; 
\vspace{-0.2cm}
\end{equation}
Each satellite has four communication terminals.

Link-capacity constraints
\begin{equation}
\label{eq_capacity}
\sum_{k} b_{k}\,x_{ij}^{k}\;\le\;C_{ij},
\qquad\forall (i,j) \text{.}\; 
\end{equation}
The sum of the bandwidths routed through any optical link must not exceed its physical capacity.

Peak CPU-power constraints
\begin{equation}
p_{i}\,z_{i}\le P_{\mathrm{lim}},
\qquad\forall i \text{.}\; 
\label{eq_power} 
\end{equation}
The peak CPU usage of satellite~$i$ should never push above the hardware limit \(P_{{lim}}\).

Link-to-node coupling
\begin{equation}
\label{eq_coupling}
x_{ij}^{k}\;\le\;y_{i}^{k},
\;\;
x_{ij}^{k}\;\le\;y_{j}^{k},
\qquad\forall(i,j),\;k \text{.}\; 
\end{equation}
A link can be used by connection $k$ only if both of its end satellites are activated for that connection.

The formulation above is a single MILP that jointly minimizes energy consumption and latency while maximizing reliability (via the negative-log penalty). Tuning the weights $\alpha$, $\beta$, and $\gamma$ moves the solution along the Pareto surface, effectively recreating the behavior of Flex-Algos 130, 129, and 128 without the need for separate optimization problems. For example, we can apply $(0,1,0)$ in high-demand scenarios to obtain optimized reliability performance.

\vspace{-0.05cm}
\section{Numerical Results}
In this work, we simulate the Lightspeed constellation and compare average CPU usage, PDR, and average end-to-end latency under various traffic scenarios. We evaluate four routing schemes—IPv6 over OSPF, plain SRv6, Green SRv6 \cite{green_srv6}, and Flex-Algo—and demonstrate that Flex-Algo achieves the best balance of energy efficiency, reliability, and low latency across all tested traffic conditions.

\subsection{Simulation Setup}

\begin{table}
\centering
\caption{Simulation parameters.}
\label{parameters}
\setlength{\tabcolsep}{12pt}
\arrayrulecolor{black}
\begin{tabular}{ll} 
\hline
\textbf{Parameter} & \textbf{Value} \\ 
\hline
Number of satellites & 198 \\
Number of ground stations & 10 \\
Number of SDN controllers & 2 \\
Max traffic capacity & 20 Mbps \\
Dynamic topology update frequency & 10 sec \\
Traffic packet timeout & 1 sec \\
($\alpha$, $\beta$, $\gamma$) for Algo 130 & (1,0,0) \\
($\alpha$, $\beta$, $\gamma$) for Algo 129 & (0,1,0) \\
($\alpha$, $\beta$, $\gamma$) for Algo 128 & (0,0,1) \\
\hline
\end{tabular}
\arrayrulecolor{black}
\vspace{-0.5cm}
\end{table}

We build our simulation on an Ubuntu virtual machine, integrating MATLAB’s Satellite Communication Toolbox to establish the Lightspeed constellation and calculate ISLs and uplink/downlink parameters including positions, distances, and propagation delays. These parameters are exported into Mininet in Linux, where the network topology updates every 10 seconds under the Ryu SDN controller. InfluxDB and Grafana capture and visualize resource metrics including CPU usage, enabling performance analysis on green TE and Flex-Algo. Inside the Ryu application, we formulate the MILP and the decision variables as link-selection bits; the weighted objective returns the best slice based on the traffic condition. When the solver returns, Ryu compares the result with the current state and routes, selects the appropriate slice, and programs an SR Policy to update the new destination SID if needed.

All simulations are conducted over 500 time slots due to the complexity of the network. Because we simulate on a PC rather than on actual satellite hardware, we use CPU usage to indicate satellite power consumption. Results therefore are machine-specific and the real satellite energy profiles would require dedicated power models. Table \ref{parameters} shows the simulation parameters used in this work. To solve the MILP, we use IBM's solver CPLEX Version 22.1.2 \cite{cplex}.

\begin{figure*}[!h]
    \centering
    \includegraphics[width=0.85\linewidth]{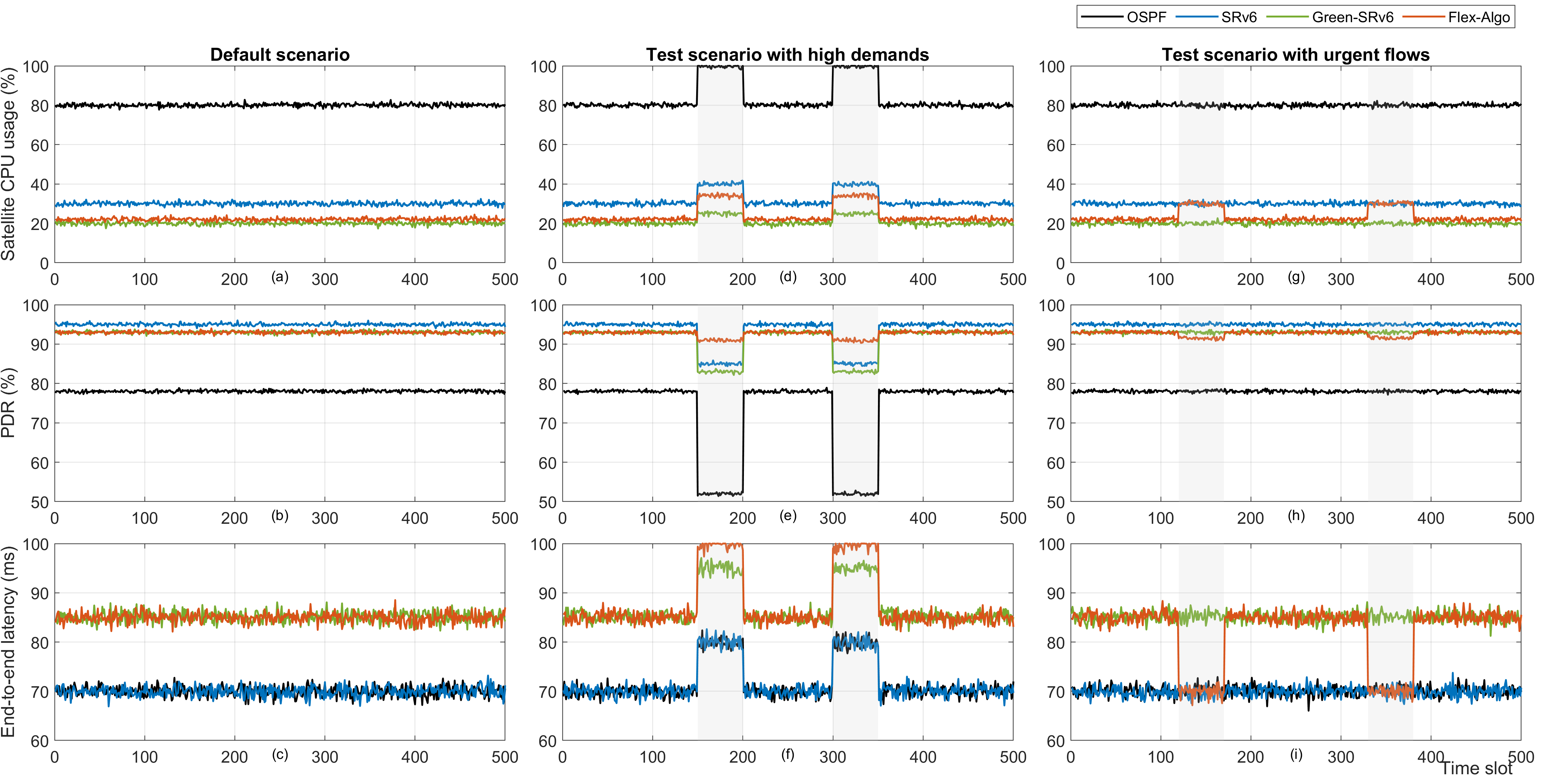}
    \caption{Average satellite CPU usage, PDR, and average end-to-end latency over time slots for three scenarios.}
    \label{result_fig}
    \vspace{-0.4cm}
\end{figure*}

\begin{figure*}[!h]
    \centering
    \includegraphics[width=0.85\linewidth]{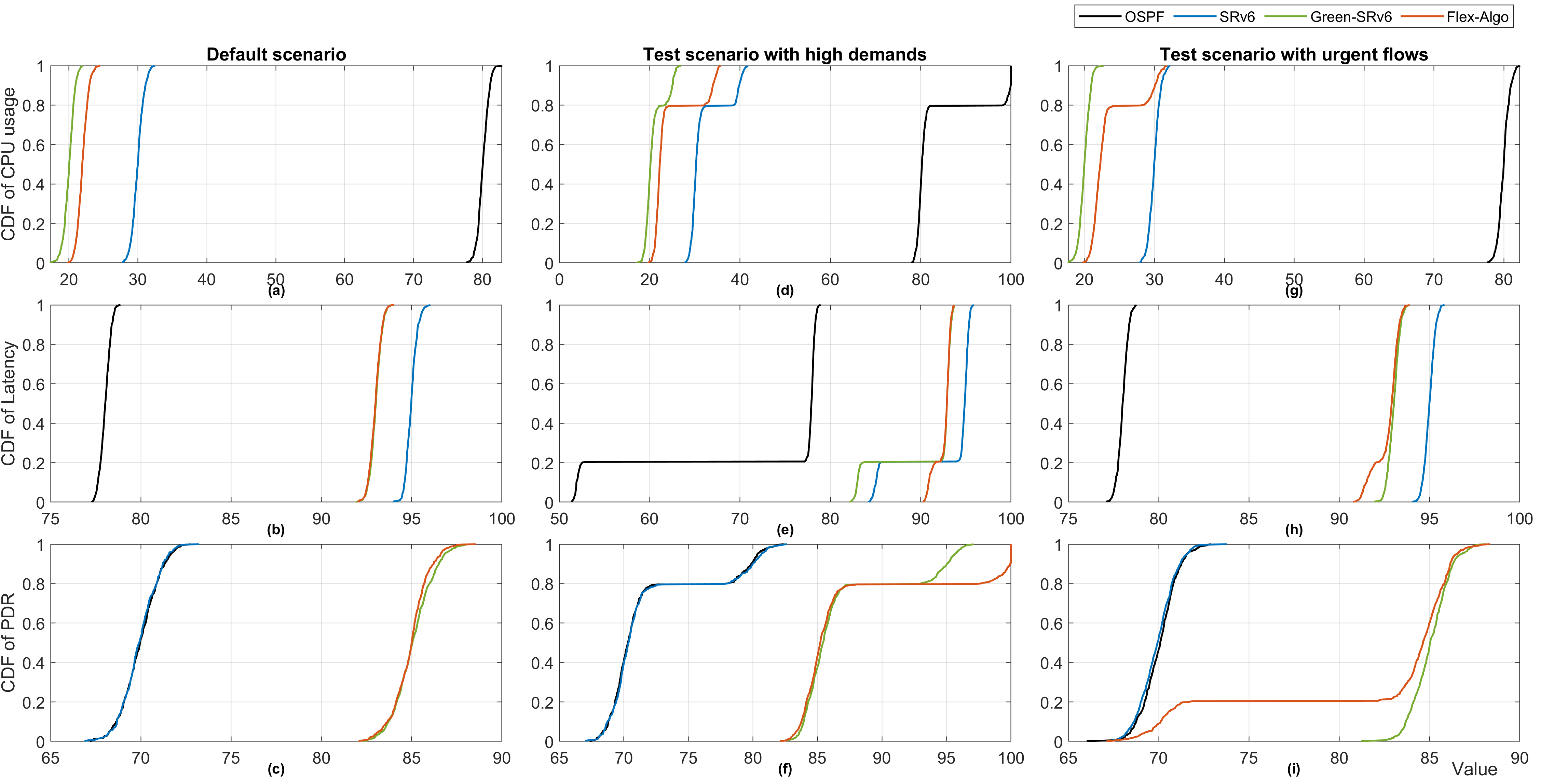}
    \caption{CDF of average satellite CPU usage, PDR, and average end-to-end latency for three scenarios.}
    \label{cdf_fig}
    \vspace{-0.5cm}
\end{figure*}

\subsection{Network Performance Results}

The numerical results for key network performance metrics, including average satellite CPU usage, PDR, and end-to-end latency over 500 time slots, are shown in Fig. \ref{result_fig} and Fig. \ref{cdf_fig}, along with the cumulative distribution functions (CDFs). The results focus on traffic exchanged between ground stations in Ottawa and Vancouver through the Lightspeed constellation. In the default scenario, the system operates in a regular green mode with a full 20 Mbps traffic load. The high-demand scenario simulates a hot spot event by increasing the load to 30 Mbps for defined intervals. Finally, the urgent-flow scenario models stock-exchange data: it tracks the delay of a single high-priority packet to evaluate latency-sensitive performance.

\vspace{-0.05cm}
Sub-figures (a), (b), and (c) show the performance of the default energy-efficient cases. CPU time series settle at four plateaus—OSPF at 80\%, SRv6 at 30\%, Green SRv6 at 20\%, and Flex-Algo at 22\%. PDRs cluster near 94\%, although OSPF drops to 78\% at full load. Latency time series separate into two bands: around 70 ms for OSPF/SRv6 and about 85 ms for the green schemes, mirrored by their CDF steps. Overall, Flex-Algo (slice 130 by default) closely matches Green SRv6: low power, high reliability, with a modest latency increase.

\vspace{-0.05cm}
Sub-figures (d), (e), and (f) show high traffic demand cases. During the gray-shaded slots, traffic demand spikes sharply. Time series show OSPF reaching 100\% CPU usage and PDR falling to 50\%. SRv6 and Green TE rise to 40\% and 25\% CPU usage, each losing about 10\% PDR and adding 10 ms latency. Flex-Algo climbs to 35\% CPU usage, slightly above the green scheme since it switches to the reliability slice Flex-Algo 129 and activates detours. This shift adds 15 ms latency, yet the CDF confirms PDR stays above 90\%. The trade-off is clear: Flex-Algo accepts a modest increase in power and delay to preserve near-lossless delivery.

\vspace{-0.05cm}
Sub-figures (g), (h), and (i) show the urgent-flow scenario. gray-shaded regions mark bursts of latency-critical packets. Only Flex-Algo reacts: CPU usage rises from 22\% to 30\%, matching SRv6, while latency falls from 85 ms to 70 ms, visible as a leftward CDF shift. The other schemes show no change, with steady end-to-end latency. PDR remains stable for the baselines and shows only a slight drop for Flex-Algo, indicating that switching to the latency-optimized slice shortens paths without significantly affecting reliability.

The paired figures highlight Flex-Algo’s adaptive benefits. Flex-Algo 130 minimizes energy under normal traffic; Flex-Algo 129 trades additional power and delay to ensure reliability during high demand; Flex-Algo 128 expends additional power to reduce latency for urgent flows. Across all scenarios, Flex-Algo delivers balanced performance and rapidly shifts operating points via policy-driven SPF slices, a real-time response not achieved by traditional OSPF or static Green SRv6. Meanwhile, the MILP runtime should also be considered and bounded to fit the control interval. In this work, the slice–selection parameters are binary and only three cases are considered, so we solve one independent MILP per slot and select the best solution according to the traffic requirement. For the Lightspeed constellation, the average runtime for the server to return after the traffic changes is around 5 seconds, which is less than one time slot to remain tractable. For larger constellations like Starlink with multiple ground stations, decomposition or hierarchical slicing can be used to reduce the reaction time and improve the scalability of the network.

\vspace{-0.1cm}
\section{Conclusion}

This work demonstrates that three SRv6 Flex-Algo slices can significantly enhance network performance in the LEO satellite constellation. By combining energy efficiency, latency, and reliability into a single MILP model and exposing each optimized topology as an IGP slice, the controller dynamically shifts traffic among Flex-Algos 130, 129, and 128 by simply updating the SR policy. Emulation results using the Lightspeed satellite constellation show that the framework matches the performance of dedicated Green SRv6 during regular traffic periods; maintains over 90\% PDR under high-demand traffic or congestion with only modest increases in power consumption and delay; and reduces urgent-flow latency to the shortest-path bound with just a slight increase in CPU usage. In contrast, legacy OSPF, SRv6, and green SRv6 each fall short on at least one metric, highlighting the benefits of slice-aware routing.

Next steps include extending the MILP to multi-commodity stochastic traffic. We will also explore reinforcement learning agents that learn slice-selection policies directly from delay-power-reliability information, potentially replacing explicit optimization at run-time, and investigate Flex-Algo coexistence with optical ISL planning and inter-operator peering in hybrid terrestrial-satellite 6G networks.

\vspace{-0.05cm}
\section{Acknowledgment}
This work has been supported by the National Research Council Canada (NRC), MDA, Mitacs, and Defence R\&D Canada (DRDC), within the Optical Satellite Communications Consortium (OSC) framework in response to the High Throughput Secure Networks (HTSN) challenge program of the Government of Canada.

\end{document}